\documentclass[11pt]{article}
\usepackage{graphicx}
\usepackage{amsmath}
\usepackage{amsfonts}
\usepackage{amssymb}
\usepackage{epsfig}
\usepackage{times}
\usepackage{cite}
\usepackage{calc}
\usepackage{version}
\usepackage[french,english]{babel}
\begin{document}

\begin{titlepage}

\null

\vskip 1.5cm
\begin{flushright}
Report: IFIC/10-15, FTUV-10-0510\\
\end{flushright}

\vskip 1.cm

{\bf\large\baselineskip 20pt
\begin{center}
\begin{Large}
Heavy quark flavour dependence of multiparticle production in QCD jets
\end{Large}
\end{center}
}
\vskip 1cm

\begin{center}

Redamy P\'erez-Ramos
\footnote{e-mail: redamy.perez@uv.es}, 
Vincent Mathieu
\footnote{e-mail: vincent.mathieu@ific.uv.es} and Miguel-Angel Sanchis-Lozano
\footnote{e-mail: miguel.angel.sanchis@ific.uv.es}\\
\medskip
Departament de F\'{\i}sica Te\`orica and IFIC, 
Universitat de Val\`encia - CSIC\\
Dr. Moliner 50, E-46100 Burjassot, Spain
\end{center}

\baselineskip=15pt

\vskip 3.5cm

{\bf Abstract}: After inserting the heavy quark mass 
dependence into QCD partonic evolution
equations, we determine the mean charged hadron
multiplicity and second multiplicity correlators of jets produced in high 
energy collisions. We thereby extend the so-called dead cone effect to the 
phenomenology of multiparticle production in QCD jets and find that the
average multiplicity of heavy-quark initiated 
jets decreases significantly as compared to the massless case, 
even taking into account
the weak decay products of the leading primary quark. 
We emphasize the relevance of our study 
as a complementary check of $b$-tagging techniques
at hadron colliders like the Tevatron and the LHC.
\end{titlepage}

\section{Introduction}

Since the very beginning of cosmic and accelerator physics, the study
of jets has been playing a prominent role in the rise and development of the 
Standard Model (SM) \cite{Leader:1996hm}. For example, the observation of 
three-jet events in electron-positron collisions at DESY
provided a direct experimental evidence of the existence of gluons.
Nowadays, QCD furnishes the theoretical framework 
for jet analysis, and conversely, jet studies furnish precise
tests of both perturbative and non-perturbative QCD, as
well as constraints and determinations of QCD parameters.    

While high-energy hadronic interactions are dominated by the production
of secondaries with rather low
transverse momentum ($p_t$) with respect to the beam axis, 
high-$p_t$ jets are expected to become 
one of the cleanest signatures for 
New Physics (NP) to be discovered at the LHC. 
On the other hand, QCD processes often are the most important 
background of such NP signatures, therefore
requiring a good understanding of QCD 
jet rates and features.  

High-$p_t$ jets can be initiated either in a short-distance
interaction among partons of the colliding hadrons, or via
electroweak (or new physics) processes. 
One well-known example
is given by the decay chain of the top quark 
$t\ \to\ H^+\ b$,  
where the $b$ quark should start a jet. Thus the ability to
identify jets from the fragmentation and hadronization
of $b$ quarks becomes very important for such Higgs boson searches.
Needless to say, the relevance of $b$-tagging extends over 
many other channels in the quest for new physics at 
hadron colliders. The experimental 
identification of $b$-jets relies upon several of their properties
in order to reject background, e.g. 
jets initiated by lighter quarks or gluons. First,
the fragmentation is hard and the leading $b$-hadron retains
a large part of the original $b$ quark momentum. In addition, 
the weak decay products may have a large transverse momentum with respect to
the jet axis therefore allowing separation from the rest of 
the cascade particles. 
Lastly, the relatively long lifetime
of $b$-hadrons leading to displaced vertices which can be
identified by using well-known impact parameter techniques \cite{Aad:2009wy}.   
Still, a fraction of light jets could be
mis-identified as $b$-jets, especially at large 
transverse momentum of the jet. Now, let us point out that an essential difference 
between heavy and light quark jets results from
kinematics constraints: the gluon radiation
off a quark of mass $m$ and energy $E\gg m$ is suppressed
inside a forward cone with an opening angle $\Theta_m=m/E$,
the so-called {\em dead cone} phenomenon \cite{Dokshitzer:1991fd,Dokshitzer:2005ri}.

In this paper, we compute the average (charged) multiplicity and multiplicity 
fluctuations of a jet initiated by a heavy quark. For this purpose,
we extend the modified leading logarithmic approximation (MLLA) 
evolution equations \cite{Dokshitzer:1991wu} to the case where the 
jet is initiated by a heavy (charm, bottom) quark. 
The average multiplicity of light quarks jets produced in high energy collisions can be 
written as $N(Y)\propto\exp\left\{\int^Y\gamma(y)dy\right\}$ ($Y$ and $y$ are
defined in section \ref{sec:kivar}), where $\gamma\simeq\gamma_0+\Delta\gamma$ 
is the anomalous dimension that accounts for
soft and collinear gluons in the double logarithmic 
approximation (DLA) $\gamma_0\simeq\sqrt{\alpha_s}$, in addition to
hard collinear gluons $\Delta\gamma\simeq\alpha_s$ or single logarithms (SLs),
which better account for energy conservation and the running of the 
coupling constant $\alpha_s$ \cite{Dokshitzer:1991wu}. Thus, 
in the MLLA both contributions can be added in the form
\begin{equation}\label{eq:gammalogic}
\gamma\simeq\sqrt{\alpha_s}+\alpha_s+{\cal O}(\alpha_s^{3/2}),
\end{equation}
such that SLs corrections with respect to 
DLA are of relative order ${\cal O}(\sqrt{\alpha_s})$. 
Notice that in jet calculus the scheme of resummation 
differs from that used in NLO 
hard scattering cross-sections calculations from the 
factorization theorem 
\cite{Curci:1980uw,Furmanski:1981cw,Collins:1998rz}, 
where no resummation over the soft gluon logarithms is performed.

As a next step in our approach, we include the {\em dead cone} 
phenomenon into the massless quark equations
by using the massive leading order (LO) splitting functions, 
first computed in QED for electrons and photons \cite{Baier:1973ms}, 
and by replacing the massless quark propagator $1/k_\perp^2$ by the massive 
one $1/(k_\perp^2+m^2)$, as it was carried out for the evaluation
of jet rates in the $e^+e^-$ annihilation
\cite{Krauss:2003cr}. Furthermore, we demonstrate that
the whole phase space of the heavy quark dipole $Q\bar Q$ 
produced in the $e^+e^-$ 
annihilation \cite{Dokshitzer:2005ri} reduces to 
that of the heavy quark jet event
in the soft and collinear limits. 

In the present work we evaluate the mean multiplicity and 
the second multiplicity correlator as a function of the mass
of the heavy quark. Actually, the mass depends on the 
scale $Q$ of the hard process 
in which the heavy quark participates. 
In the $\overline{MS}$ renormalization scheme, for example,  
the running mass becomes a function of the
strong coupling constant $\alpha_s(Q)$,
and the pole mass defined through 
the renormalized
heavy quark propagator (for a review see \cite{Kluth:2006bw} and references therein). 
Nonetheless, in our QCD leading-order analysis no distinction between 
running and pole mass of heavy quarks has to be made.
On the other hand, the quark mass considered 
as a parameter in our calculations
should not play a significant role in the jet evolution 
until the virtuality $Q$ drops down to values of the order of the
quark mass. Therefore, the dominance of the dead cone phenomenon
at low $Q$ justifies neglecting the running of the 
heavy quark mass, hence adopting a value close to the pole mass
as a natural cut-off for final-state soft and collinear singularities.
In order to assess the scheme-dependence of
our calculations beyond leading-order
we have considered a broad range of possible values for 
the charm and bottom masses:   
$m_c=1-1.5$ GeV and $m_b=3-5$ GeV, respectively \cite{Kluth:2006bw}.

Lastly, we will see that under the assumption of local parton 
hadron duality (LPHD)
as hadronization model \cite{Azimov:1984np,Dokshitzer:1995ev}, 
light- and heavy-quark initiated jets show significant differences
regarding particle multiplicities as a consequence of soft gluon suppression inside
the dead cone. Such differences could be exploited
by using auxiliary criteria complementing $b$-tagging procedures to
be applied to jets with very large transverse momentum, as advocated
in this paper.

\section{Kinematics and variables}
\label{sec:kivar}
As known from jet calculus for light quarks, 
the evolution time parameter determining the structure
of the parton branching of the primary gluon is given by 
(for a review see \cite{Dokshitzer:1991wu} and references therein)
\begin{equation}\label{eq:masslessev}
y=\ln\left(\frac{k_\perp}{Q_0}\right),\quad k_\perp=zQ\geq Q_0,
\quad Q=E\Theta\geq Q_0,
\end{equation}
where $k_\perp$ is the transverse momentum of the gluon emitted off the
light quark, $Q$ is the virtuality of the jet (or jet hardness), 
$E$ the energy of the 
leading parton, $Q_0/E\leq\Theta\leq\Theta_0$  
is the emission angle of the gluon ($\Theta\ll1$), $\Theta_0$ the 
total half opening angle of the jet being 
fixed by experimental requirements, 
and $Q_0$ is the collinear cut-off parameter.
Let us define in this context the variable $Y$ as
\begin{equation}
y=Y+\ln z,\quad Y=\ln\left(\frac{Q}{Q_0}\right).
\end{equation}
The appearance of this scale is a consequence of angular ordering
(AO) of successive
parton branchings in QCD cascades \cite{Azimov:1984np,Dokshitzer:1991wu}. 
An important difference in the structure of light ($\ell\equiv q=u,d,s$) 
versus heavy quark ($h\equiv Q=c,b$) jets stems from the dynamical 
restriction on the phase space of primary gluon radiation in the heavy quark 
case, where the gluon radiation off an energetic quark $Q$ with mass
$m$ and energy $E\gg m$ is suppressed inside the forward cone with an
opening angle $\Theta_m=m/E$, the above-mentioned
dead cone phenomenon \cite{Dokshitzer:2005ri}. This effect is in
close analogy to QED, where photon radiation is also suppressed
at small angles with respect to a moving massive charged particle
(e.g. tau versus muon).

The corresponding evolution time parameter for a jet initiated 
by a heavy quark with energy $E$ and mass $m$ appears in a natural 
way and reads \cite{Dokshitzer:2005ri}
\begin{equation}
\tilde y=\ln\left(\frac{\kappa_\perp}{Q_0}\right),\quad 
\kappa_\perp^2=k_\perp^2+z^2m^2,
\end{equation}
which for collinear emissions $\Theta\ll1$ can also be rewritten in the form
\begin{equation}\label{eq:Qtheta}
\kappa_\perp=z\tilde Q,\quad \tilde Q=E\left(\Theta^2+\Theta_m^2\right)^{\frac12},
\end{equation}
with $\Theta\geq\Theta_m$ (see Fig.\,\ref{fig:Qsplit}). 

An additional comment is in order concerning the 
AO for gluons emitted off the heavy quark. In (\ref{eq:Qtheta}), $\Theta$ is the  
emission angle of the primary gluon $g$ being emitted off the heavy quark. Now
let $\Theta'$ be the emission angle of a second gluon $g'$ relative to the primary 
gluon with energy $\omega'\ll\omega$ and $\Theta''$ the emission angle relative to 
the heavy quark; in this case 
the {\em incoherence} condition $\Theta'^2\leq(\Theta^2+\Theta_m^2)$ 
(see appendix \ref{app:AO}) together with 
$\Theta''>\Theta_m$ (the emission angle of the second gluon should still be larger 
than the dead cone) naturally leads (\ref{eq:Qtheta}) 
to become the proper evolution
parameter for the gluon subjet (for more details see \cite{Dokshitzer:2005ri}). 
For $\Theta_m=0$, the standard 
AO ($\Theta'\leq\Theta$) is recovered.
Therefore, for a massless quark, the virtuality of the
jet simply reduces to $Q=E\Theta$ as given above.
The same quantity $\kappa_\perp$ determines the scale of the running coupling
$\alpha_s$ in the gluon emission off the heavy quark. It can be related to the anomalous
dimension of the process by
\begin{equation}
\gamma_0^2(\kappa_\perp)=
2N_c\frac{\alpha_s(\kappa_\perp)}{\pi}=\frac1{\beta_0(\tilde y+\lambda)},\quad
\beta_0(n_f)=\frac1{4N_c}\left(\frac{11}3N_c-\frac23n_f\right),\quad 
\lambda=\ln\frac{Q_0}{\Lambda_{QCD}},
\end{equation}
where $n_f$ is the number of active flavours and $N_c$ the number of colours.
The variation of the effective 
coupling $\alpha_s$ as $n_f\to n_f+1$ over the heavy quarks threshold has 
been suggested by next-to-leading (NLO) calculations in the $\overline{MS}$ scheme 
\cite{Dokshitzer:1995ev} as well. However, since the inclusion of NLO terms to the coupling constant
provides corrections of relative order ${\cal{O}}(\alpha_s)$ and therefore beyond
${\cal{O}}(\sqrt{\alpha_s})$, this effect is subleading in the MLLA (see (\ref{eq:gammalogic})). 
In this context $\beta_0(n_f)$ will be evaluated at the total number of quarks we consider in our application. The four scales of the process are related as follows,
$$
\tilde Q\gg m\gg Q_0\sim\Lambda_{QCD},
$$
where $Q_0\sim\Lambda_{QCD}$ corresponds to the limiting spectrum 
approximation \cite{Dokshitzer:1991wu}. Finally, the dead cone 
phenomenon imposes the following bounds of integration to the perturbative regime
\begin{equation}\label{eq:bounds}
\frac{m}{\tilde Q}\leq z\leq 1-\frac{m}{\tilde Q},\quad
m^2\leq\tilde Q^2\leq E^2(\Theta_0^2+\Theta_m^2),
\end{equation} 
which now account for the phase-space of the heavy quark jet. The last inequality
states that the minimal transverse momentum of the jet $\tilde Q=E\Theta_m=m$ 
is given by the mass of the heavy quark, which enters the game as the natural 
cut-off parameter of the perturbative approach.
\section{Definitions and notation}
\label{sec:defandnot}
The multiplicity distribution is defined by the formula
\begin{equation}
P_n=\frac{\sigma_n}{\sum_{n=0}^{\infty}\sigma_n}
=\frac{\sigma_n}{\sigma_{inel}},\quad \sum_{n=0}^{\infty}P_n=1
\end{equation}
where $\sigma_n$ denotes the cross section of an $n$-particle yield process, 
$\sigma_{inel}$ is the inelastic cross-section, and the sum runs 
over all possible values of $n$.

It is often more convenient to represent multiplicity distributions by 
their moments. All such sets can be 
obtained from the generating functional
$Z(y,u)$ \cite{Dokshitzer:1991wu} defined by 
$$
Z(y,u)=\sum_{n=0}^{\infty}P_n(y)\ (1+u)^n
$$
at the energy scale $y$. For fixed $y$, we can drop this variable 
from the {\em azimuthally} 
averaged generating functional $Z(u)$; the moments are then 
calculated from the MLLA master equation as
$$
F_q=\frac{1}{\left<n\right>^q}\frac{d^qZ(u)}{du^q}\Big|_{u=0},\quad
K_q=\frac{1}{\left<n\right>^q}\frac{d^q\ln Z(u)}{du^q}\Big|_{u=0},\quad
C_q=\frac{1}{\left<n\right>^q}\frac{d^qZ(e^u-1)}{du^q}\Big|_{u=0}
$$
where the average multiplicity is defined by the formula,
$$
\left<n\right>\equiv N=\sum_{n=0}^{\infty}P_nn,\quad
P_n=\frac{1}{n!}\frac{d^nZ(u)}{du^n}\Big|_{u=-1}. 
$$
$F_q$ are respectively the factorial moments, 
often called multiplicity correlators, 
$K_q$ are the cumulants of rank $q$ and $C_q$, the moments of the multiplicity 
distribution $P_n$. The multiplicity correlator and the moment of rank $q=2$
are related as follows,
$$
F_2=\frac{\left<N(N-1)\right>}{N^2}=C_2-N^{-1}.
$$ 
Moreover, the width $D^2=\left<N^2\right>-N^2$ of the multiplicity 
distribution $P_n$ can be written in the equivalent forms,
$$
D^2=(F_2-1)N^2+N=K_2N^2+N=(C_2-1)N^2.
$$
In terms of Feynman diagrams, $F_q$ correspond to the set of all graphs 
while $K_q$ describe the connected diagrams. Therefore, $K_q$ are more suited 
for the construction of the evolution equations.

Specifically, we will compute the average multiplicity 
of partons in jets to be denoted hereafter as $N_A$, with $A=Q,q,g$, 
corresponding to a heavy, light quark or gluon initiated jet
respectively. Likewise, we will compute 
the second rank multiplicity correlator inside the same jet.

Once arrived at his point, let us make an important distinction
between two different particle sources populating
heavy-quark initiated jets. 
On the one hand, parton cascade from gluon emission
yields the QCD component of the total jet multiplicity
(the main object of our present study), {\em excluding weak
decay products of the leading primary quark}
at the final stage of hadronization. On the other hand,
the latter products coming from the leading flavoured hadron
should be taken into account in the measured multiplicities
of jets. We shall denote the average charged hadron multiplicity
from the latter source as $N_A^{dc}$. Hence
the total charged average multiplicity, $N_A^{total}$, reads
\begin{equation}\label{eq:total}
N_A^{total}=N^{ch}_A+N_A^{dc}\ ;\quad A=q,Q.
\end{equation} 
As a consequence of the LPHD, $N^{ch}_A={\cal K}^{ch}\times N_A$ 
\cite{Dokshitzer:1995ev,Azimov:1984np}, where
the free parameter ${\cal K}^{ch}$ normalizes the average multiplicity
of partons to the average multiplicity of charged hadrons.
For charm and bottom quarks, we will respectively set the values
$N_c^{dc}=2.60 \pm 0.15$ and $N_b^{dc}=5.55 \pm 0.09$
\cite{Dokshitzer:2005ri,Akers:1995ww}, while in light quark jets one expects
$N_q^{dc}=1.2 \pm 0.1$ \cite{:1994qa}.

Now let us point out the distinct trends 
from each contribution to (\ref{eq:total}) as the quark mass
increases. The dead cone effect suppresses $N_Q$
for heavier quark masses. Conversely, $N_Q^{dc}$ 
becomes more significant for bottom jets.
As we shall later see, the former will ultimately
dominate the behaviour of the total average multiplicity $N_Q^{total}$
of heavy quark jets for high $Q$ values. In this paper, we advocate
the use of such a difference between average jet multiplicities as   
a signature to distinguish {\em a posteriori}
heavy from light quark jets, particularly in $b$-tagging techniques
applied to the analysis of many interesting decay channels.  

\section{QCD evolution equations}
The splitting functions \cite{Ellis:1991qj}
\begin{equation}
P(z,\alpha_s)=\alpha_sP^{(0)}(z)+\alpha_s^2P^{(1)}(z)+\ldots
\end{equation}
where $P^{(0)}(z)$ and $P^{(1)}(z)$ are respectively the LO and NLO 
splitting functions, can be associated to each vertex of 
the process in the partonic shower. $P(z,\alpha_s)$
determines the decay probability of a parent parton (quark, anti-quark, gluon) into
two offspring partons of energy fractions $z$ and $1-z$. In this paper, we are rather 
concerned with calculations which only involve the LO $P^{(0)}(z)$ 
splitting functions in the evolution equations \cite{Dokshitzer:1991wu}. 
At LO or tree level, the Lorentz structure
of the massless and massive splitting functions is universal for the  
decays $e^{\pm}\to e^{\pm}\gamma$, $\gamma\to e^+e^-$ and $q\to qg$, $g\to q\bar q$ 
in QED and QCD respectively \cite{Peskin:1995ev,Baier:1973ms}. 
The inclusion of the NLO splitting functions 
\footnote{The dependence of our approach 
on a certain renormalization scheme would arise with the inclusion 
of the NLO QCD splitting functions in the evolution.} in our 
approach would provide corrections of relative order ${\cal O}(\alpha_s)$, 
which going beyond ${\cal O}(\sqrt{\alpha_s})$ in (\ref{eq:gammalogic}), 
are subleading in this frame.
\begin{figure}[h]
\begin{center}
\epsfig{file=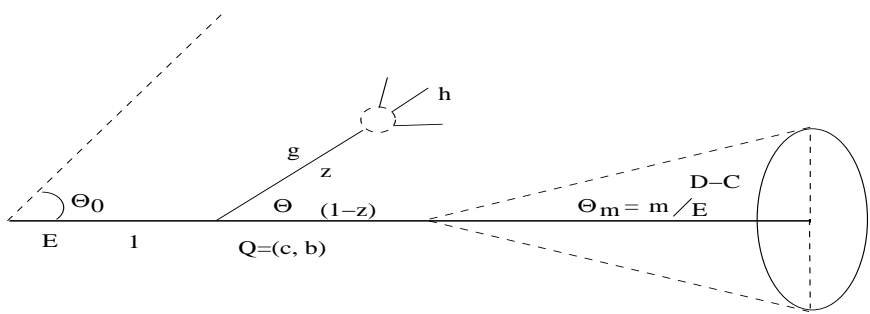, height=4truecm,width=11.5truecm}
\caption{\label{fig:Qsplit} Parton splitting in the process $Q\to Qg$: 
a {\em dead cone} with opening angle $\Theta_m$
is schematically shown.}  
\end{center}
\end{figure}

Let us start by considering the
the splitting process, $Q\to Qg$, $Q$ being a heavy quark and $g$ 
the emitted gluon which is displayed in Fig.\ref{fig:Qsplit}; the corresponding LO
splitting function reads \cite{Baier:1973ms,Krauss:2003cr}
\begin{equation}\label{eq:PQg}
P_{Qg}^{(0)}(z)=
\frac{C_F}{N_c}\left[\frac{1}{z}-1+\frac{z}2-
\frac{z(1-z)m^2}{k_\perp^2 + z^2m^2}\right],\quad
P_{QQ}^{(0)}(z)=P_{Qg}^{(0)}(1-z)
\end{equation}
where $k_\perp\approx min(zE\Theta,(1-z)E\Theta)$ is the transverse 
momentum of the soft gluon being
emitted off the heavy quark. The previous formula (\ref{eq:PQg}) has the 
following physical interpretation, for $k_\perp\ll z^2m^2$, the corresponding
limit reads $P_{Qg}(z)\to\frac{C_F}{2N_c}z$ and that is why, at leading logarithmic
approximation, the forward emission of soft and collinear gluons off the heavy 
quark becomes suppressed once $\Theta\ll\Theta_m$, while the
emission of hard and collinear gluons dominates in this region.

For the massless process $g\to gg$, we adopt the standard three gluon 
vertex kernel \cite{Dokshitzer:1991wu,Dremin:2000ep}
\begin{equation}\label{eq:Pgg}
P^{(0)}_{gg}(z)=\frac{1}{z}-(1-z)[2-z(1-z)],
\end{equation}
and finally for $g\to Q\bar Q$, we take \cite{Baier:1973ms,Krauss:2003cr}
\begin{equation}\label{eq:PgQ}
P_{gQ}^{(0)}(z)=\frac1{4N_c}\left[1-2z(1-z)+\frac{2z(1-z)m^2}{k_\perp^2+m^2}\right],
\end{equation}
which needs to be resummed together with the three gluon vertex contribution. 
However, as a first approach to this problem, we neglect the production
of heavy quark pairs inside gluon and quark jets, making use of 
\cite{Dokshitzer:1991wu,Dremin:2000ep}
\begin{equation}
P^{(0)}_{gq}(z)=P^{(0)}_{gQ}(z)|_{m=0}=\frac1{4N_c}\left[1-2z(1-z)\right].
\end{equation}
Including mass effects in the evolution equations also requires the replacement
of the massless quark propagator $1/k_\perp^2$ by the massive quark propagator
$1/(k_\perp^2+z^2m^2)$ \cite{Krauss:2003cr}, 
such that the phase space for soft and collinear 
gluon emissions off the heavy quark can be written 
in the form \cite{Krauss:2003cr}
\begin{equation}\label{eq:jetcs}
d^2\sigma_{Qg}\simeq
\gamma_0^2
\frac{dk_\perp^2}{k_\perp^2+z^2m^2}
dzP_{Qg}^{(0)}(z),
\end{equation}
where $P_{Qg}^{(0)}(z)$ is given by (\ref{eq:PQg}). Working out the structure
of (\ref{eq:PQg}) and setting $k_\perp\approx zE\Theta$ one has
\begin{equation}\label{eq:PQgbis}
P_{Qg}^{(0)}(z)=
\frac{C_F}{N_c}\left[\frac{1}{z}\frac{\Theta^2}{\Theta^2+\Theta_m^2}-1+\frac{z}2
+\frac{\Theta_m^2}{\Theta^2+\Theta_m^2}\right],
\end{equation}
such that, one can recover the phase space for the
soft and collinear gluon emission in the DLA
\begin{equation}\label{eq:deadcone}
d^2\sigma_{Qg}\simeq\frac{C_F}{N_c}
\gamma_0^2
\frac{\Theta^2d\Theta^2}{(\Theta^2+\Theta_m^2)^2}
\frac{dz}{z}.
\end{equation}
Notice that the overall multiplicity of the process $e^+e^-\to Q\bar Qg$ cannot be represented 
simply as the sum of three independent parton multiplicities \cite{Dokshitzer:2005ri}. 
As stressed in \cite{Dokshitzer:1991wu}, the accompanying multiplicity off
the quark dipole becomes dependent on the geometry of the whole 
jet ensemble in a Lorentz invariant way and should be treated as a different problem. 
However, as suggested in \cite{Dokshitzer:1991wu}
and demonstrated in the appendix 
\ref{app:dipole}, we recover the correct limit of the one 
jet event (\ref{eq:jetcs},\ref{eq:PQgbis}) from the heavy quark dipole 
\cite{Dokshitzer:2005ri}, in the soft $z\ll1$ and collinear 
$\Theta\ll1$ gluon limits.

According to the Low-Burnett-Kroll theorem \cite{Low:1958sn,Burnett:1967km}, the $dz/z$
part of the radiation density has a classical origin and is, therefore,
universal, independent of the intrinsic quantum numbers and the process,
while the other terms are quantum corrections.
The system of two-coupled evolution equations for the gluon and quark jets
average multiplicity in the massless case at MLLA  
simplifies to the following in the hard splitting 
region $k_\perp\sim E\Theta$ ($z\sim1-z\sim1$) 
\cite{Dremin:2000ep,Capella:1999ms}
\begin{eqnarray}
\frac{d}{dY}N_q(Y)\!&\!=\!&\!\int_{Q_0/Q}^{1-Q_0/Q} dz\,
\gamma_0^2(z)P^{(0)}_{qg}(z)N_g(Y+\ln z)\label{eq:NQhbis},\\\notag\\
\frac{d}{dY}N_g(Y)\!&\!=\!&\!
\int_{Q_0/Q}^{1-Q_0/Q} dz\,\gamma_0^2(z)\left[P^{(0)}_{gg}(z)N_g(Y+\ln z)
+n_fP^{(0)}_{gq}(z)
\left(2N_q(Y)-N_g(Y)\right)\right],\label{eq:NGhbis}
\end{eqnarray}
where $P^{(0)}_{qg}(z)=P^{(0)}_{Qg}(z)|_{m=0}$ and $P^{(0)}_{gq}(z)=P^{(0)}_{gQ}(z)|_{m=0}$ according to
(\ref{eq:PQg}) and (\ref{eq:PgQ}) respectively. It is obtained from the MLLA master equation for
the azimuthally averaged generating functional $Z(y,u)$, by taking the functional 
derivative over $u$ (see section \ref{sec:defandnot}).
The arguments of $N_A$ in the right hand side of the 
equations do not depend on $z$ because for hard partons $z\sim1$, the original
arguments $Y+\ln z$ and $Y+\ln(1-z)$ of these functions can be 
approximated to $Y$ after $\ln z$ and $\ln(1-z)$ are neglected.
Substituting (\ref{eq:PQgbis}) into (\ref{eq:NQhbis}), replacing
$E\Theta$ by $E(\Theta^2+\Theta_m^2)^{1/2}$ in the argument of all logs, which is 
equivalent to replacing the massless propagator by the massive one and the argument
of the running coupling $k_\perp=zE\Theta$ by $\kappa_\perp=zE(\Theta^2+\Theta_m^2)^{1/2}$,
after taking the bounds (\ref{eq:bounds}) and integrating over the regular part of the 
splitting function (\ref{eq:PQgbis}), one has the equation
\begin{equation}\label{eq:massprop}
\frac{N_c}{C_F}\frac{dN_Q}{d\tilde Y}=\frac{\Theta^2}{\Theta^2+\Theta_m^2}
\int_{\tilde Y_{m}}^{\tilde Y_{ev}}dy\gamma_0^2(\tilde y)N_g(\tilde y)-\left(\frac34-
\frac{3\Theta_m}{2(\Theta^2+\Theta_m^2)^{\frac12}}-\frac{\Theta_m^2}
{\Theta^2+\Theta_m^2}\right)\gamma_0^2(\tilde Y)N_g(\tilde Y),
\end{equation}
where
$$
\tilde Y_{m}=L_m,\quad 
\tilde Y_{ev}\approx\tilde Y,
$$
finally, introducing the chain of transformations
$$
\frac{\Theta^2}{\Theta^2+\Theta_m^2}
=1-\frac{\Theta_m^2}{\Theta^2+\Theta_m^2}=
1-e^{-2\tilde Y+2L_m},\quad L_m=\ln\frac{m}{Q_0},
$$
where the $L_m$ logarithms are new in this
context and provide power suppressed corrections 
to the solution of
the evolution equations. Since massless, in the gluon jet however, the evolution 
time variable remains the same (\ref{eq:masslessev}) as that in the massless quark jet. 
Nevertheless, the argument of the average 
multiplicity in the gluon subjet $N_g(\tilde y)$ in (\ref{eq:massprop}), 
depends on the same argument $\tilde y$. That is why, in the following, we insert 
the mass of the heavy quark in the argument of all logs 
in (\ref{eq:NGhbis}) and take the same integration bounds (\ref{eq:bounds}) 
correspondingly. 
The system of QCD evolution equations now reads
\begin{eqnarray}
\frac{N_c}{C_F}\frac{dN_Q}{d\tilde Y}\!\!&\!\!=\!\!&\!\!
\epsilon_1(\tilde Y,L_m)\int_{\tilde Y_{m}}^{\tilde Y_{ev}}
d\tilde y\gamma_0^2(\tilde y)N_g(\tilde y)-\epsilon_2(\tilde Y,L_m)\gamma_0^2(\tilde Y)N_g(\tilde Y),
\label{eq:NQhter}\\
\frac{dN_g}{d\tilde Y}\!\!&\!\!=\!\!&\!\!
\int_{\tilde Y_{m}}^{\tilde Y_{ev}}d\tilde y\gamma_0^2(\tilde y)N_g(\tilde y)-A(\tilde Y,L_m)
\gamma_0^2(\tilde Y)N_g(\tilde Y),
\label{eq:NGhter}
\end{eqnarray}
where
\begin{equation}\label{eq:Anf}
A(\tilde Y,L_m)=a(n_f)-\left[2+\frac{n_f}{2N_c}
\left(1-2\frac{C_F}{N_c}\right)\right]e^{-\tilde Y+L_m}+
\frac12\left[1+\frac{n_f}{N_c}\left(1-2\frac{C_F}{N_c}\right)\right]e^{-2\tilde Y+2L_m},
\end{equation}
with
\begin{equation}
a(n_f)=\frac{1}{4N_c}\left[\frac{11}3N_c+\frac23n_f
\left(1-2\frac{C_F}{N_c}\right)\right]
\end{equation}
and
\begin{equation}\label{eq:epsilon12}
\epsilon_1(\tilde Y,L_m)=1-e^{-2\tilde Y+2L_m},\quad 
\epsilon_2(\tilde Y,L_m)=\frac34-\frac{3}{2}
e^{-\tilde Y+L_m}-e^{-2\tilde Y+2L_m}.
\end{equation}
There are the following kinds of power suppressed corrections to the heavy quark multiplicity: 
the leading integral term of
(\ref{eq:NQhter}) is ${\cal O}(\frac{m^2}{\tilde Q^2})$ suppressed, while subleading MLLA 
corrections appear in the standard form ${\cal O}(\sqrt{\alpha_s})$ like in the massless case, finally
${\cal O}(\frac{m}{\tilde Q}\sqrt{\alpha_s})$ and
${\cal O}(\frac{m^2}{\tilde Q^2}\sqrt{\alpha_s})$, which are new in this context. Similar corrections 
have been found in the treatment of multiparticle production off the heavy quark dipole 
\cite{Dokshitzer:1991wu} and in the computation of the heavy quark content inside gluon 
jets \cite{Mueller:1985zp}.  

Similar power suppressed corrections proportional 
to $e^{-\tilde Y+L_m}$ in $A(\tilde Y, L_m)$  
and $\epsilon_2(\tilde Y,L_m)$
were reported in \cite{Capella:1999ms} for the massless case. Indeed, 
such results can be recovered after setting $m/\tilde Q\to Q_0/Q$ in 
(\ref{eq:Anf}) and (\ref{eq:epsilon12}). 
For massive particles however, these terms are somewhat larger
and can not be neglected in our approach unless they are evaluated for much 
higher energies than at present colliders. 
On top of that, the corresponding massless equations in the high energy limit are obtained 
from (\ref{eq:NQhter}) and (\ref{eq:NGhter}) simply by setting $\tilde y\to y$, $\tilde Y\to Y$, 
$Y_{ev}\to Y$, $Y_{m}\to 0$, $\epsilon_1\to1$,
$$
\epsilon_2\to\tilde\epsilon_2=\frac34-\frac32e^{-Y}+{\cal O}\left(e^{-2Y}\right),
\quad A\to\tilde A=a(n_f)-\left[2+\frac{n_f}{2N_c}
\left(1-2\frac{C_F}{N_c}\right)\right]e^{-Y}+{\cal O}\left(e^{-2Y}\right),
$$
and are written in the standard form \cite{Dremin:2000ep,Capella:1999ms}
\begin{eqnarray}\label{eq:Nqlight}
\frac{N_c}{C_F}\frac{dN_q}{dY}\!\!&\!\!=\!\!&\!\!
\int_{0}^{Y}
d y\gamma_0^2(y)N_g(y)-\tilde\epsilon_2(Y)\gamma_0^2(Y)N_g(Y),\\
\frac{dN_g}{dY}\!\!&\!\!=\!\!&\!\!
\int_{0}^{Y}dy\gamma_0^2(y)N_g(y)-\tilde A(Y)
\gamma_0^2(Y)N_g(Y),
\label{eq:Nglight}
\end{eqnarray} 
with the initial condition $N_{g,q}(Y=0)=1$ at threshold.
Notice that (\ref{eq:NQhter}) and (\ref{eq:NGhter}) are valid only for $m\gg Q_0$ and
therefore $m\to 0$ does not reproduce the correct limit, which has to be smooth as given 
by the massless equations (\ref{eq:Nqlight}) and (\ref{eq:Nglight}).

As can be seen from (\ref{eq:NQhter}), 
the function $\epsilon_1$ 
also gives the power suppressed 
contribution $\propto -e^{-2\tilde Y+2L_m}$ which decreases the production
of soft and collinear gluons off the heavy quark, however, this contribution
is power suppressed ${\cal O}(\frac{m^2}{\tilde Q^2})$ and turns out to be
rather small as the energy scale increases. Since heavy quarks are less sensitive to recoil effects, 
the subtraction terms $\propto e^{-\tilde Y+L_m}$ and $\propto e^{-2\tilde Y+2L_m}$ 
in $\epsilon_2(\tilde \tilde Y,L_m)$ diminish
the role of energy conservation as compared to massless quark jets.
As a consistency check, upon integration over $\tilde Y$ of
the DLA term in Eq.(\ref{eq:NQhter}), the phase space structure of 
the radiated quanta
in (\ref{eq:deadcone}) is recovered:

\begin{equation}\label{eq:dconesimple}
N_Q(\ln\tilde Q)\approx1+\frac{C_F}{N_c}\int_{0}^{\Theta^2_0}
\frac{\Theta^2d\Theta^2}{(\Theta^2+\Theta_m^2)^2}\int_{m/\tilde Q}^{1-{m/\tilde Q}}\frac{dz}{z}
\left[\gamma_0^2N_g\right](\ln z\tilde Q).
\end{equation}
Notice that the lower bound over $\Theta^2$ in (\ref{eq:dconesimple}) ($\tilde Y$ in (\ref{eq:NQhter}))
can be taken down to ``0" ($Y_m=L_m$ in (\ref{eq:NQhter})) because the heavy quark mass plays the 
role of collinear cut-off parameter.

\subsection{Towards the solution of the evolution equations}
\begin{figure}[h]
\begin{center}
\epsfig{file=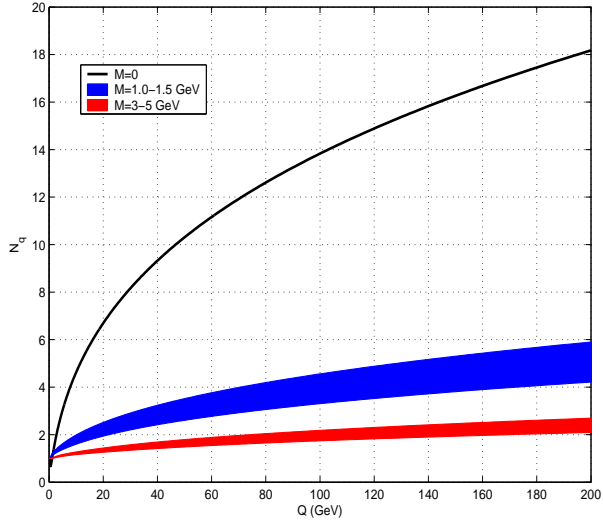, height=7.0truecm,width=8.0truecm}
\caption{\label{fig:NQ} Massless 
and massive quark jet average multiplicity $N_Q$
as a function of the jet hardness $Q$. Bands indicate $m_c$ and $m_b$ in the $[1,1.5]$ and 
$[3,5]$ GeV intervals respectively.}  
\end{center}
\end{figure}

First we solve the self-contained equation for the gluon 
jet (\ref{eq:NGhter}). The second and third exponential
terms in (\ref{eq:Anf}) are slowly varying functions of the variable $\tilde Y$ 
at high energy scales.
In the same limit and for the sake of simplicity we set
the bound of integration over $(z,\tilde y)$ to 
$\tilde Y_{ev}\to\tilde Y$, $\tilde Y_{m}\to0$ and solve the 
equation by performing the Mellin transform
\begin{equation}\label{eq:Mellin}
N_g(\tilde Y)=\int_C\frac{d\omega}{2\pi i}e^{\omega\tilde Y}\tilde N_g(\omega),
\end{equation} 
where the contour $C$ lies to the right of all singularities in the 
complex plane of $\omega$. Replacing (\ref{eq:Mellin}) into (\ref{eq:NGhter}) leads the
first order differential equation in Mellin space:
\begin{equation}
\frac{d\tilde N_g}{d\omega}=\left(\lambda\omega-\frac1{\beta_0\omega}-A\right)\tilde N_g,\quad
A\stackrel{\tilde Y\gg1}=A(\tilde Y,L_m),
\end{equation}
which upon inversion leads to the following solution
\begin{equation}
N_g(\tilde Y,L_m)=const\times (\tilde Y+\lambda)^{-\Sigma}
\exp\left(2\sqrt{\frac{\tilde Y+\lambda}{\beta_0}}\right).
\end{equation}
Then, the initial condition at threshold, which is 
reached when the jet virtuality approaches the mass of the heavy quark yields
\begin{equation}
N_g(L_m)=1\Longrightarrow const=(\tilde Y_{m}+\lambda)^{\Sigma}
\exp\left(-2\sqrt{\frac{\tilde Y_{m}+\lambda}{\beta_0}}\right)
\end{equation}
and finally,
\begin{equation}\label{eq:Ngsol}
N_g(\tilde Y,L_m)\approx 
\left(\frac{\tilde Y+\lambda}{\tilde Y_{m}+\lambda}\right)^{-\Sigma}
\exp\left(2\sqrt{\frac{\tilde Y+\lambda}{\beta_0}}
-2\sqrt{\frac{\tilde Y_{m}+\lambda}{\beta_0}}\right),
\end{equation}
with
$$
\Sigma\stackrel{\tilde Y\gg1}=\Sigma(\tilde Y,L_m)=\frac{A(\tilde Y,L_m)}{2}-\frac{\beta_0}{4}.
$$
From (\ref{eq:Ngsol}) and by making use of $N_g(\tilde Y)\simeq\exp{\left(\int^{\tilde Y} d\tilde y 
\gamma(\tilde y)\right)}$ \cite{Dokshitzer:1991wu}, one gets the rate of multiplicity growth 
as a function of $\tilde Y$ to be,
$$
\gamma\approx \gamma_0-\Sigma\gamma_0^2.
$$
A similar solution without power corrections, which was written for the 
``gluon mass" was given in \cite{Mueller:1985zp}. 
Notice that the fact of introducing the ``gluon mass" in this context is 
technical rather than physical. However, phenomenological observations favour
a dynamically generated mass for the gluon \cite{Mathieu:2008me}.
In order to obtain the approximate 
solution of (\ref{eq:NQhter}), as before we consider
the functions 
$$\epsilon_1\stackrel{\tilde Y\gg1}=\epsilon_1(\tilde Y,L_m),
\quad \epsilon_2\stackrel{\tilde Y\gg1}=\epsilon_2(\tilde Y,L_m)
$$ 
as constants at high energy scale. Subtracting (\ref{eq:NQhter}) from (\ref{eq:NGhter}) and setting 
$$
\frac{N_c}{C_F}
\frac{d}{d\tilde Y}\left(\epsilon_1^{-1}\frac{dN_Q}{d\tilde Y}\right)=
\gamma_0^2N_g(\tilde Y)
$$
on the right hand side of the outcoming, one has
\begin{equation}\label{eq:subst1}
\frac{dN_g}{d\tilde Y}-\frac{N_c}{C_F}\epsilon_1^{-1}\frac{dN_Q}{d\tilde Y}=
-\frac{N_c}{C_F}\left[A
-\epsilon_1^{-1}\epsilon_2\right]
\frac{d}{d\tilde Y}\left(\epsilon_1^{-1}\frac{dN_Q}{d\tilde Y}\right)
\end{equation}
Working out the structure of (\ref{eq:subst1}), so as to obtain the ratio
\begin{equation}
r=\frac{N_g}{N_Q},
\end{equation}
we can rewrite it in the form
\begin{eqnarray}
\frac{dN_g}{d\tilde Y}-\frac{N_c}{C_F}\frac{dN_Q}{d\tilde Y}+
\frac{N_c}{C_F}\left(1-\epsilon_1^{-1}\right)\frac{dN_Q}{d\tilde Y}=
-\frac{N_c}{C_F}\left(A-\epsilon_1^{-1}\epsilon_2\right)
\frac{d}{d\tilde Y}\left(\epsilon_1^{-1}\frac{dN_Q}{d\tilde Y}\right).
\end{eqnarray}
Finally, we obtain
\begin{equation}\label{eq:ratio}
r(\tilde Y,L_m)\stackrel{\tilde Y\gg1}\approx\frac{N_c}{C_F}\epsilon_1^{-1}(\tilde Y,L_m)\left[1-\left(A(\tilde Y,L_m)-
\epsilon_1^{-1}(\tilde Y,L_m)\epsilon_2(\tilde Y,L_m)\right)\gamma_0\right],
\end{equation}
which becomes valid in the limit
\begin{equation}
\frac{dr_1}{d\tilde Y}\stackrel{\tilde Y\gg1}\approx0,
\quad r_1(\tilde Y,L_m)\stackrel{\tilde Y\gg1}\approx\left(A(\tilde Y,L_m)-
\epsilon_1^{-1}(\tilde Y,L_m)\epsilon_2(\tilde Y,L_m)\right).
\end{equation}
Finally, the approximate
average multiplicity in a jet initiated by a heavy quark reads
\begin{equation}\label{eq:NHeavyQ}
N_Q(\tilde Y,L_m)=\frac{C_F}{N_c}\epsilon_1(\tilde Y,L_m)
\frac{N_g(\tilde Y,L_m)}{1-r_1(\tilde Y,L_m)\gamma_0},
\end{equation} 
where $\epsilon_1(\tilde Y,L_m)$ is written in (\ref{eq:epsilon12}) and 
$N_g(\tilde Y,L_m)$ is given by (\ref{eq:Ngsol}). Thus, as the mass of the leading heavy quark 
increases, the multiparticle yield in the heavy quark jet is affected by 
power corrections, by the suppression of the anomalous dimension 
$\gamma_0=\gamma_0(m^2)$ and mainly by the massive suppressed exponential contribution arising
from the initial condition at threshold. However, for the sake of completeness,
we solve the evolution equations numerically and display the energy dependence
of the average multiplicity in Fig.\ref{fig:NQ}. The asymptotic behaviour 
of the distribution is then seen to follow the expected exponential increase
given by (\ref{eq:NHeavyQ}), with $N_g$ in (\ref{eq:Ngsol}). 
Finally, we estimate
the difference between the light and heavy quark jet multiplicities, which yields,
\begin{equation}\label{eq:Nqdiff}
N_q-N_Q\stackrel{E\to\infty}{\approx}\left[1-\exp\left(-2\sqrt{\frac{L_m}{\beta_0}}\right)\right]N_q,\quad
N_q\propto\exp2\sqrt{\frac{\tilde Y}{\beta_0}}.
\end{equation}
Hence,  (\ref{eq:Nqdiff}) is exponentially increasing because it is dominated by
the leading DLA energy dependence of $N_q$. According to (\ref{eq:Nqdiff}), the gap 
arising from the dead cone effect should be bigger for the $b$ than for the $c$ quark
at the primary state bremsstrahlung radiation off the heavy quark jet.
The approximated solution of the 
evolution equations leads to the rough behaviour of $N_q-N_Q$ in 
(\ref{eq:Nqdiff}), which is not exact in its present form. In Fig.\ref{fig:NQ}, we display
the numerical solution of the evolution equations (\ref{eq:Nqlight}) for $N_q$ and
(\ref{eq:NQhter}) for $N_Q$ corresponding to the heavy quark mass intervals 
$m_c=1-1.5$ GeV, $m_b=3-5$ GeV.
Let us remark that the gap arising between the light quark jet multiplicity and
the heavy quark jet multiplicity follows the trends given
by (\ref{eq:Nqdiff}) asymptotically with $E\to\infty$. In particular, notice that
the dispersion of the mean multiplicities becomes irrelevant for the purposes of our study. 
This behaviour should not be confused with that followed by the $Q\bar Q$ antenna in the $e^+e^-$ 
annihilation, where the difference is roughly constant and energy independent 
\cite{Dokshitzer:2005ri,Kisselev:2008fw}. Indeed, (\ref{eq:Nqdiff}) can not be 
extrapolated to the dipole case by simply setting $N_{Q,\bar Q}\approx 2N_Q$ because the evolution
equations do not take into account 
interference effects between the $Q$ and the $\bar Q$ jets in the $e^+e^-$ annihilation.
Finally, as expected for massless quarks $L_m=0$, the  difference $N_q-N_Q$ vanishes.

\section{Heavy quark evolution of second multiplicity correlator}

The second multiplicity correlator 
was first considered for massless quarks in \cite{Malaza:1985jd}. 
It is defined in the form $N_A^{(2)}=\langle {N_A}({N_A}-1)\rangle$
in gluon ($A=g$) and quark ($A=q$) jets. 
The normalized second multiplicity
correlator defines the width of the multiplicity distribution and is related to its 
dispersion squared $D_A^2=\langle N_A^2\rangle-N_A^2$ 
by the formula (see definitions and notation in section \ref{sec:defandnot})
\begin{equation}
D_A^2=(F_{A,2}-1)N_A^2+N_A.
\end{equation}
The second multiplicity correlators normalized to their own 
average multiplicity squared are
\begin{equation}\label{eq:G2Q2}
F_{2,g}\equiv G_2=\frac{\langle {N_g}({N_g}-1)\rangle}{N_g^2},\quad 
F_{2,q}\equiv Q_2=\frac{\langle {N_q}({N_q}-1)\rangle}{N_q^2},
\end{equation}
inside a gluon and a quark jet respectively.
These observables are obtained by integrating the double differential inclusive 
cross section over the energy fractions $x_1=e_1/E$ and $x_2=e_2/E$ of two particles
emitted inside the jet,
$$
\langle {N_A}({N_A}-1)\rangle=\iint dx_1dx_2\left(\frac1{\sigma}
\frac{d^2\sigma}{dx_1dx_2}\right)_A.
$$
The system of evolution equations for light quarks following from
the MLLA master equation can be written as \cite{Dokshitzer:1991wu,Dremin:2000ep},

\vbox{
\begin{eqnarray}
\frac{d}{dY}(N_q^{(2)}-N_q^2)\!\!&\!\!=\!\!&\!\!\int_{Q_0/Q}^{1-Q_0/Q}dz\gamma_0^2(z)
P^{(0)}_{qg}(z) N_g^{(2)}(Y+\ln z)\!\label{eq:corrq}\,,\\\cr
\frac{d}{dY}(N_g^{(2)}-N_g^2)\!\!&\!\!=\!\!&\!\!\int_{Q_0/Q}^{1-Q_0/Q}dz
\gamma_0^2(z)P^{(0)}_{gg}(z)N_g^{(2)}(Y+\ln z)\cr\cr
\!\!&\!\!+\!\!&\!\!n_f\int_{Q_0/Q}^{1-Q_0/Q}dz\gamma_0^2(z)P^{(0)}_{gq}(z)
\left[2\Big(N_q^{(2)}(Y)\!-\!N_q^2(Y)\Big)
\!-\!\Big(N_g^{(2)}(Y)\!-\!
N_g^2(Y)\Big)\right.\cr\cr
\!\!&\!\!+\!\!&\!\!\left.
\Big(2N_q(Y)-N_g(Y)\Big)^2\right]\label{eq:corrg},
\end{eqnarray}
}
with the following relations at DLA \cite{Dokshitzer:1982xr,Dokshitzer:1982ia},
\begin{equation}
N_q^{(2)}-N_q^2=\frac{C_F}{N_c}\left(N_g^{(2)}-N_g^2\right),\quad N_q=\frac{C_F}{N_c}N_g.
\end{equation}
The arguments of $N_A^{(2)}$ and $N_A^2$ in the right hand side of the 
equations do not depend on $z$ because for hard partons $z\sim1$, the original
arguments $Y+\ln z$ and $Y+\ln(1-z)$ of these functions can be 
approximated to $Y$ after $\ln z$ and $\ln(1-z)$ are neglected.
Substituting (\ref{eq:PQgbis}) into (\ref{eq:corrq}), after replacing
$E\Theta$ by $E(\Theta^2+\Theta_m^2)^{1/2}$ in the argument of all logs, 
taking the bounds (\ref{eq:bounds}) and 
integrating over the regular part of the splitting functions, one 
has the system 
\begin{eqnarray}
\frac{N_c}{C_F}\frac{d}{d\tilde Y}(N_Q^{(2)}-N_Q^2)\!\!&\!\!=\!\!&\!\!
\epsilon_1(\tilde Y,L_m)\int_{\tilde Y_{m}}^{\tilde Y_{ev}}
d\tilde y\gamma_0^2(\tilde y)N_g^{(2)}(\tilde y)-
\epsilon_2(\tilde Y,L_m)\gamma_0^2(\tilde Y)N_g^{(2)}(\tilde Y),
\label{eq:N2Qh}\\
\frac{d}{d\tilde Y}(N_g^{(2)}-N_g^2)\!\!&\!\!=\!\!&\!\!
\int_{\tilde Y_{m}}^{\tilde Y_{ev}}d\tilde y\gamma_0^2(\tilde Y)N_g^{(2)}(\tilde y)-A(\tilde Y,L_m)
\gamma_0^2(\tilde Y)N_g^{(2)}(\tilde Y)\cr
\!\!&\!\!+\!\!&\!\!\Big(A(\tilde Y,L_m)-B(\tilde Y,L_m)\Big)\gamma_0^2(\tilde Y)N_g^2(\tilde Y),
\label{eq:N2Gh}
\end{eqnarray}
with the initial conditions $N_A^{(2)}(L_m)=\frac{dN_A^{(2)}}{d\tilde Y}(L_m)=0$, where 
\begin{equation}
B(\tilde Y,L_m)=b(n_f)-\left[2-\frac{n_f}{2N_c}\left(1-2\frac{C_F}{N_c}\right)^2\right]e^{-\tilde Y+L_m}
+\frac12\left[1-\frac{n_f}{N_c}\left(1-2\frac{C_F}{N_c}\right)^2\right]e^{-2\tilde Y+2L_m},
\end{equation}
with
$$
b(n_f)=\frac1{4N_c}\left[\frac{11}{3}N_c-\frac{2}{3}n_f\left(1-2\frac{C_F}{N_c}\right)^2\right].
$$
Accordingly, in the massless limit, (\ref{eq:N2Qh}) and (\ref{eq:N2Gh}) reduce to 
\cite{Malaza:1985jd,Dremin:1999ub}
\begin{eqnarray}
\frac{N_c}{C_F}\frac{d}{dY}(N_Q^{(2)}-N_Q^2)\!\!&\!\!=\!\!&\!\!
\int_{0}^{Y}
dy\gamma_0^2(y)N_g^{(2)}(y)-
\tilde\epsilon_2(Y)\gamma_0^2(Y)N_g^{(2)}(Y),
\label{eq:N2Qhm0}\\
\frac{d}{dY}(N_g^{(2)}-N_g^2)\!\!&\!\!=\!\!&\!\!
\int_{0}^{Y}dy\gamma_0^2(y)N_g^{(2)}(y)-\tilde A(Y)
\gamma_0^2(Y)N_g^{(2)}(Y)\cr
\!\!&\!\!+\!\!&\!\!\Big(\tilde A(Y)-\tilde B(Y)\Big)\gamma_0^2(Y)N_g^2(Y),
\label{eq:N2Ghm0}
\end{eqnarray}
where
$$
\tilde B(Y)=b(n_f)-\left[2-\frac{n_f}{2N_c}\left(1-2\frac{C_F}{N_c}\right)^2\right]e^{-Y}
+{\cal O}\left(e^{-2Y}\right).
$$
The functions $\tilde\epsilon_2(Y)$ and $\tilde A(Y)$ are defined above through the 
equations for the average multiplicity (\ref{eq:Nqlight}) and (\ref{eq:Nglight}) in light quark jets.
\subsection{Approximate solution of the evolution equations}
For the gluon jet, taking into account that at high energy scales one has 
$A\stackrel{\tilde Y\gg1}{\approx}A(\tilde Y,L_m)$ and
$B\stackrel{\tilde Y\gg1}{\approx}B(\tilde Y,L_m)$ 
($\frac{dA,B}{d\tilde Y}\stackrel{\tilde Y\gg1}{\approx}0$), 
and making use of (\ref{eq:Ngsol}), the solution reads \cite{Ramos:2008as}
\begin{equation}\label{eq:G2}
G_2(\tilde Y,L_m)-1\stackrel{\tilde Y\gg1}{\approx}\frac13-C_1(\tilde Y,L_m)\gamma_0,
\end{equation}
where, 
$$
C_1(\tilde Y,L_m)\stackrel{\tilde Y\gg1}{\approx}-
\frac29A(\tilde Y,L_m)+\frac19\beta_0(n_f)+\frac23B(\tilde Y,L_m),
\quad \frac{dC_1}{d\tilde Y}\stackrel{\tilde Y\gg1}{\approx}0. 
$$
Accordingly, for the quark jet one finds \cite{Ramos:2008as}

\begin{equation}\label{eq:Q2}
Q_2(\tilde Y,L_m)-1\stackrel{\tilde Y\gg1}{\approx}\frac{N_c}{C_F}
\epsilon_1^{-1}(\tilde Y,L_m)
\left(\frac13-\tilde C_1(\tilde Y,L_m)\gamma_0\right),
\end{equation}
where
$$
\tilde C_1(\tilde Y,L_m)\stackrel{\tilde Y\gg1}{\approx}\frac5{18}A(\tilde Y,L_m)+\frac19\beta_0(n_f)+\frac16B(\tilde Y,L_m),
\quad \frac{d\tilde C_1}{d\tilde Y}\stackrel{\tilde Y\gg1}{\approx}0.
$$
Therefore, the correlators $G_2$ (\ref{eq:G2}) and $Q_2$ (\ref{eq:Q2}) are mainly affected by
power corrections ${\cal O}(\frac{m}{\tilde Q}\sqrt{\alpha_s})$ 
and ${\cal O}(\frac{m^2}{\tilde Q^2}\sqrt{\alpha_s})$ which diminish the role of energy conservation 
in a heavy quark jet and make the correlation stronger as the particle yield gets suppressed 
inside the dead cone region. Thus, with such effects, the correlators increase as the mass
of the leading heavy quark increases and approach the asymptotic DLA values $G_2=\frac43$ and $Q_2=1+\frac{N_c}{3C_F}$ respectively. However, for realistic energy scales 
this approximation fails, in particular because of the integration over the 
dead cone term $\propto\epsilon_1(\tilde Y,L_m)$
in the double logarithmic contribution of (\ref{eq:N2Qh}). That is why, as
we further emphasize in the appendix \ref{sec:analitnumer1},
we should rather display the numerical solution
of the equations (\ref{eq:N2Qh}) and (\ref{eq:N2Gh}) in the relevant energy range.

\section{Phenomenological consequences}

The study of multiplicity distributions (mean and higher rank momenta)
and inclusive correlations
has been traditionally employed in the analysis of multiparticle
production in high energy hadron collisions, notably regarding soft
(low $p_t$) physics (see e.g. \cite{Dremin:2000ep} and references therein). 
Moreover, the use
of inclusive particle correlations 
has been recently advocated in the search of 
new phenomena \cite{SanchisLozano:2008te}.

On the other hand, it is well known that the study of average charged hadron 
multiplicities of jets 
in $e^+e^-$ collisions
has also become a useful tool 
for testing (perturbative) QCD calculations 
(see \cite{Dokshitzer:2005ri,Kisselev:2008fw} and references
therein).

\begin{figure}[h]
\begin{center}
\epsfig{file=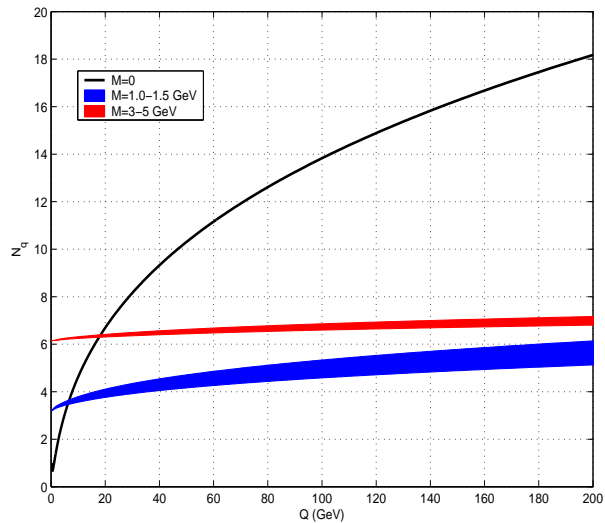, height=7truecm,width=8truecm}
\caption{\label{fig:NQdecays} Massless
and massive quark jet average multiplicity $N_Q^{total}$
as a function of the jet hardness $Q$ including
heavy quark flavour decays. Same comments as in Fig.\,\ref{fig:NQ}.}  
\end{center}
\end{figure}

In this paper we advocate the role of 
mean multiplicities of jets as a
potentially useful signature of new physics when
combined with other selection criteria.
In Fig.\,\ref{fig:NQdecays}, we plot as function of the jet 
hardness $Q$\footnote{The energy range $100\leq Q (\text{GeV})\leq200$
should be realistic for Tevatron and LHC phenomenology.}, 
the total average jet multiplicity (\ref{eq:total}), which accounts for
the primary state radiation off the heavy quark 
together with the decay products from the final-state flavoured hadrons, which
were introduced in section \ref{sec:defandnot}. 
For these predictions, we set ${\cal K}^{ch}=0.6$ in (\ref{eq:total}),
which we take from \cite{:2008ec} and $Q_0\sim\Lambda_{QCD}=230$ MeV \cite{Dokshitzer:1991wu}.
Moreover, the flavour decays 
constants $N_c^{dc}=2.60 \pm 0.15$ and $N_b^{dc}=5.55 \pm 0.09$ are independent
of the hard process inside the cascade, such that $N_A^{dc}$ can be added in the 
whole energy range. For instance, such values were obtained by the OPAL
collaboration at the $Z^0$ peak of the $e^+e^-$ annihilation.
In this experiment, $D^*$ mesons were properly reconstructed in order to provide
samples of events with varying $c$ and $b$ purity. By studying the charged 
hadron multiplicity in conjunction with samples of varying $b$ purity, it became possible
to measure light and heavy quark charged hadron multiplicities
separately \cite{Akers:1995ww}.
As compared to the average multiplicities of the primary 
state radiation displayed in Fig.\,\ref{fig:NQ}, after accounting for $N_A^{dc}$, the $b$ quark jet 
multiplicity becomes slightly higher than the $c$ quark jet multiplicity, although
both remain suppressed because of the dead cone effect. 
\begin{figure}[h]
\begin{center}
\epsfig{file=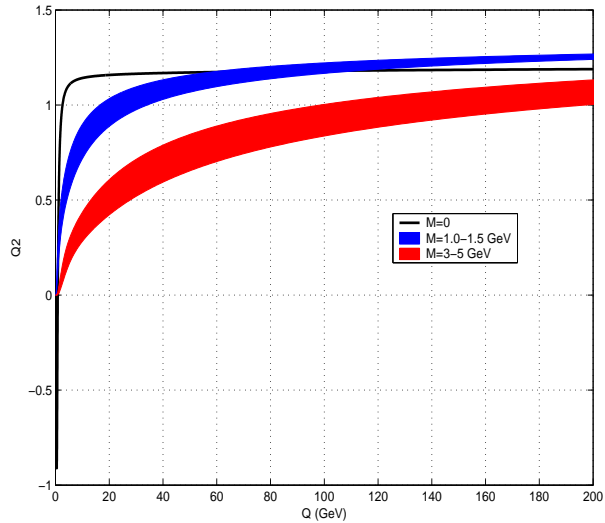, height=7truecm,width=8truecm}
\caption{\label{fig:NQ2} Massless 
and massive quark jet correlator $Q_2$
as a function of the jet hardness $Q$. Same comments as in Fig.\,\ref{fig:NQ}.}  
\end{center}
\end{figure}

The second quark jet correlator defined in (\ref{eq:G2Q2}) 
for different flavours is displayed in Fig.\,\ref{fig:NQ2} as a function
of the jet hardness $Q$. Since contributions to the dispersion 
from quark flavour decays are negligible ($N_c^{dc}=2.60 \pm 0.15$ and $N_b^{dc}=5.55 \pm 0.09$)
the correlation is the strongest for partons at the primary state radiation 
of the process. Notice that, while the $u,d,s$ and the
$c$ quark correlators are of the same order of magnitude for a jet hardness $Q\gtrsim40$ GeV as
relevant energy range, the vertical difference with the $b$ quark correlator, 
which is weaker, can still exceed $\sim20\%$. Therefore, the measurement of the quark 
correlator should provide a further signature of $b$ flavour and associated exotic particles 
yield when compared with $u,d,s,c$
correlators. Finally, the variation of the charged hadrons 
average multiplicities and the correlator in the above-mentioned intervals 
for the charm and and bottom masses turns out to be negligible for our purposes.

\section{Conclusions}

Jet physics has been so far of paramount
importance in the rise and development of the SM and expectedly
will keep such a prominent role in the discovery of new phenomena
at hadron colliders like the Tevatron and the LHC. However, QCD
jets represent a formidable challenge to disentangle
signals of new physics from hadronic background in most
cases. On the other hand, plenty of
new physics channels end with heavy flavours in the final state,
before fragmenting and hadronizing. 

Thus, our present work focusing on the differences of the average charged
hadron multiplicity between jets initiated by gluons, light or heavy quarks
could indeed represent a helpful auxiliary criterion to tag 
such heavy flavours from background for jet hardness $Q \gtrsim 40$ GeV.
Notice that we are suggesting as a potential signature
the {\em a posteriori} comparison
between average jet multiplicities corresponding to 
different samples of events where other criteria
to discriminate heavy from light quark initiated jets
were first applied. In other words, one should compare
mean multiplicities at different jet-hardness $Q$, 
in order to check that they agree with QCD predictions. 
Fig.\ref{fig:NQdecays}  plainly demonstrate that the separation between light quark
jets and heavy quark jets is allowed above a few tens
of GeV with the foreseen errors of the experimentally measured average 
multiplicities of jets. The difference between light quark jet multiplicities 
and heavy quark jet multiplicities $N_q-N_Q$ in one jet is exponentially increasing because
of suppression of forward gluons in the angular region around the heavy quark direction. This
result is not drastically affected after accounting for heavy flavour decays multiplicities,
such that it can still be used as an important signature for the search 
of new physics in a jet together
with other selection criteria. However, our result can only be applied to single jets
and therefore, it should not be extrapolated to the phenomenology of the $Q\bar Q$ dipole 
treated in \cite{Dokshitzer:2005ri} because neither interference effects
with other jets nor large angle gluon emissions are considered in our case. 
As a complementary observable, in particular for $b$-tagging, the 
second multiplicity correlator (\ref{eq:G2Q2}) displayed in Fig.\,\ref{fig:NQ2} 
should also contribute to discriminate $b$ quark 
from $u,d,s,c$ quark channels. Indeed, while the $c$ quark correlator remains 
of the same order of magnitude than the light quark jet correlator, the $b$ quark
correlator gets weaker by 20$\%$ and therefore, distinguishable with respect to
the other quarks in the relevant energy range. Furthermore, the inclusion of the
heavy quark mass in the evolution equations for the 
correlator does not affect the asymptotic energy independent flattening of the slope 
arising from the KNO (Koba-Nielsen-Olsen) scaling \cite{Koba:1972ng}.

Notice that the measurement of such observables require the previous reconstruction
of jets at hadron colliders. Thanks to important recent
developments on jet reconstruction algorithms \cite{Cacciari:2005hq,Cacciari:2008gn,Cacciari:2007fd}, future analysis such as single
inclusive hadron production inside light and
heavy quark jets look very promising.

\section*{Acknowledgements}

We gratefully acknowledge interesting discussions with F. Arleo, J.P Guillet,
E. Pilon, G. Rodrigo, S. Sapeta,  M. Vos and
C. Troestler for helping us with numerical recipes.
R.P.R acknowledges support from Generalitat Valenciana under grant PROMETEO/2008/004
and M.A.S, from FPA2008-02878 and GVPROMETEO2010-056. V.M acklowledges support from
the grant HadronPhysics2, a FP7-Integrating 
Activities and Infrastructure Program of the
European Commission under Grant 227431, by UE (Feder) and the MICINN 
(Spain) grant FPA2007-65748-C02- and by GVPrometeo2009/129.

\appendix

\section{From AO to the incoherence condition of gluon emission off the heavy quark}
In the MLLA, the parton decay probabilities are written in a form 
\cite{Dokshitzer:1991wu},
\label{app:AO}

\begin{equation}\label{eq:cswV}
d^2_{Qg}\equiv d^2\sigma_{Q\to Qg}=\frac{\alpha_s}{2\pi}P^{(0)}_{Qg}(z)dzV(\vec n)\frac{d\Omega}{8\pi},
\quad V_{g(Q)}^{g'}(\vec n)=\frac{a_{g'Q}+a_{gQ}-a_{g'gf}}{a_{g'g}a_{g'Q}}.
\end{equation}

\begin{figure}[h]
\begin{center}
\epsfig{file=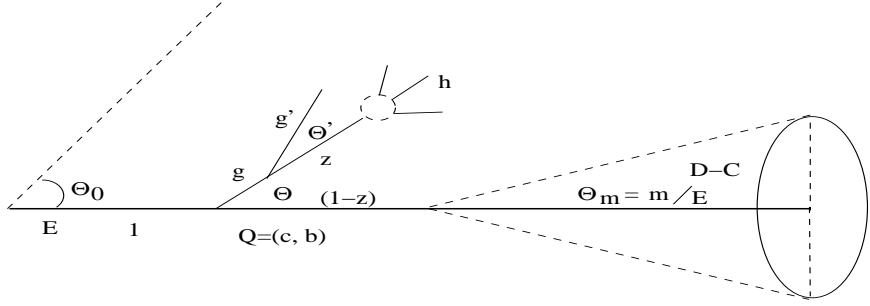, height=4truecm,width=11.5truecm}
\caption{\label{fig:Qsplitbis} Second gluon emission off the primary gluon
in the process $Q\to \bar Q+g(f)+g'(s)$.}  
\end{center}
\end{figure}

It describes the process $Q\to \bar Q+g(f)+g'(s)$ displayed in Fig.\ref{fig:Qsplitbis}, 
where the subscripts mean father and son.
In this case we define $\Theta=\Theta_{gQ}$, $\Theta'=\Theta_{g'g}$. For light quarks involved
in the same process $q\to \bar q+g(f)+g'(s)$, if ``i" and ``k" denote the
massless particles, then the angular factor $a_{ik}$ 
in the relativistic case is written as
\begin{equation}
a_{ik}=1-\cos\Theta_{ik}.
\end{equation}
After taking the azimuthal average around 
the ``son" gluon direction, one obtains 
\cite{Dokshitzer:1991wu},
\begin{equation}\label{eq:Vfqs}
\left<V_{g(q)}^{g'}\right>=\int_{0}^{2\pi}\frac{d\phi}{2\pi}V_{g(q)}^{g'}(\vec n)=
\frac2{a_{g'g}}\vartheta(a_{gq}-a_{g'g}),
\end{equation}
with $\vartheta$ the Heaviside function. This leads to the exact AO inside partonic 
cascades by replacing the strong AO in the DLA $\Theta'\ll\Theta$ by $\Theta'\leq\Theta$
in the MLLA. For massive particles we may write (\ref{eq:cswV}) in the same form after 
replacing the standard massless splitting functions \cite{Dokshitzer:1991wu} by the 
massive one \cite{Krauss:2003cr}. If the leading parton is a heavy quark,
the angular factor of the emitted gluon $``g"$ off the heavy quark $Q$, can be 
checked after some simple kinematics, to be written in the form,
\begin{equation}
a_{gQ}=1-\sqrt{1-\Theta_{m}^2}\cos\Theta,
\end{equation}
where $\Theta_m$ is the angle of the dead cone. In this case, (\ref{eq:Vfqs}) can be 
rewritten as follows
\begin{equation}\label{eq:Vfqsbis}
\left<V_{g(Q)}^{g'}\right>=\int_{0}^{2\pi}\frac{d\phi}{2\pi}V_{g(Q)}^s(\vec n)=
\frac2{a_{g'g}}\vartheta(a_{gQ}-a_{g'g}),
\end{equation}
imposing that $\cos\Theta'\geq\sqrt{1-\Theta_{m}^2}\cos\Theta$. For small angles, if one sets
$\cos\Theta\approx1-\frac{\Theta^2}2$ in both members of the previous inequality, one gets the incoherent
condition:
\begin{equation}\label{eq:incohcond}
\Theta'^2\leq\Theta^2+\Theta_m^2.
\end{equation}
In the massless case $\Theta_m=0$, (\ref{eq:incohcond}) simply reduces to the 
standard exact AO $\Theta'\leq\Theta$.

\section{Accompanying radiated quanta off the heavy quark dipole}
\label{app:dipole}

In \cite{Dokshitzer:2005ri,Dokshitzer:1995ev}, the probability of soft gluon emission
of the heavy quark pair $Q\bar Q$ produced in the $e^+e^-$ 
annihilation was written in the form,
\begin{equation}\label{eq:QQbarr}
d^2\sigma_{Q\bar Qg}=\frac{C_F\alpha_s(\kappa_\perp^2)}{\pi}\frac{dz}{z}
\frac{\beta}{v}d\cos\Theta_c\left\{2(1-z)\frac{\beta^2\sin^2\Theta_c}{(1-\beta^2\cos^2\Theta_c)^2}
+z^2\left[\frac1{1-\beta^2\cos^2\Theta_c}-\frac12\right]\zeta_V^{-1}\right\},
\end{equation}
where $z$ is the energy fraction of the emitted gluon and $\Theta_c$ the emission angle with respect
to the center of mass of the $Q\bar Q$ pair. Moreover, the following notation was introduced:
\begin{eqnarray}
\beta^2=1-\frac{4m^2}{W^2(1-z)}, \quad v^2=1-\frac{4m^2}{W^2},\quad \zeta_V=1+2\frac{m^2}{W^2}.
\end{eqnarray}
The transverse momentum of the gluon appearing on the argument of the running coupling in (\ref{eq:QQbarr})
was written in the form,
\begin{equation}\label{eq:kappaperp}
\kappa_\perp^2=\left(\frac{zW}2\right)^2(1-\beta^2\cos^2\Theta_c)^2.
\end{equation}
With such a notation, we now take interest in the soft ($1-z\sim1$) and collinear ($\Theta_c\ll1$) 
limits of (\ref{eq:QQbarr}) and set $W\to2E$ in order to reduce (\ref{eq:QQbarr}) to the single
jet event initiated by a heavy quark $Q$. Thus, the terms take the following form
\begin{itemize}
\item
$$
\beta^2\stackrel{z\ll1}{\approx}1-\frac{m^2}{E^2}=1-\Theta_m^2,\quad (1-\beta^2\cos^2\Theta_c)^2
\stackrel{z\ll1,\Theta_c\ll1}{\approx}(\Theta_c^2+\Theta_m^2)^2,
$$
\item
$$
\frac{2(1-z)\beta^2\sin^2\Theta_c}{(1-\beta^2\cos^2\Theta_c)^2}\stackrel{z\ll1,\Theta_c\ll1}{\approx}
\frac{2(1-z)\Theta_c^2}{(\Theta_c^2+\Theta_m^2)^2},\;\; \left[\frac{z^2}{1-\beta^2\cos^2\Theta_c}-\frac12\right]
\zeta_V^{-1}\stackrel{z\ll1,\Theta_c\ll1}{\approx}\frac{z^2}{\Theta_c^2+\Theta_m^2}.
$$
\end{itemize}
The term proportional to $-\frac12$ in the cross section (\ref{eq:QQbarr}) does contribute neither as a 
soft logarithmic contribution nor as a collinear one, and therefore
can be neglected in this 
approximation. It should correspond to a Feynman diagram which 
only accounts for interference effects between
the $Q$ and the $\bar Q$ jets in the $Q\bar Q$ antenna. 
In this limit, for one jet we set $d^2\sigma_{Q\bar Qg}\to d^2\sigma_{Qg}$, 
$\Theta_c\to\Theta$ and $\frac{\beta}{v}\to1$, such that the 
dipole cross section (\ref{eq:QQbarr}) can be rewritten in the form
\begin{eqnarray}\label{eq:dsigmaQg}
d^2\sigma_{Qg}&\simeq&\frac{C_F}{N_c}\gamma_0^2(\kappa_\perp^2)\frac{d\Theta^2}{\Theta^2+\Theta_m^2}
\left\{\frac{1}{z}\frac{\Theta^2}{\Theta^2+\Theta_m^2}-1+\frac{z}2
+\frac{\Theta_m^2}{\Theta^2+\Theta_m^2}\right\}dz\\
&\equiv&\gamma_0^2(\kappa_\perp^2)\frac{d\kappa_\perp^2}{\kappa_\perp^2}P^{(0)}_{Qg}(z)dz,\notag
\end{eqnarray}
as given in (\ref{eq:jetcs}), where $P^{(0)}_{Qg}(z)$ was written in (\ref{eq:PQgbis}), 
while (\ref{eq:kappaperp}) was reduced to the following,
$$
\kappa_\perp^2\stackrel{z\ll1,\Theta_c\ll1}{\approx}z^2E^2(\Theta^2+\Theta_m^2)\equiv z^2\tilde Q^2,\quad 
\tilde Q^2=E^2(\Theta^2+\Theta_m^2).
$$
Therefore, in the soft and collinear approximation, the dipole case (\ref{eq:QQbarr}) 
reduces to the jet event (\ref{eq:dsigmaQg}), which coincides with the expression given in
(\ref{eq:jetcs}). In the massless case $\Theta_m=0$, it simplifies to the standard DGLAP kernel
as explained in \cite{Dokshitzer:2005ri},
$$
d^2\sigma_{Qg}\propto\frac{d\Theta^2}{\Theta^2}\left\{2(1-z)+z^2\right\}\frac{dz}{z}.
$$

\section{Analytical versus numerical solution of the heavy quark correlator
equation (\ref{eq:N2Qh})}

\label{sec:analitnumer1}

\begin{figure}[h]
\begin{center}
\epsfig{file=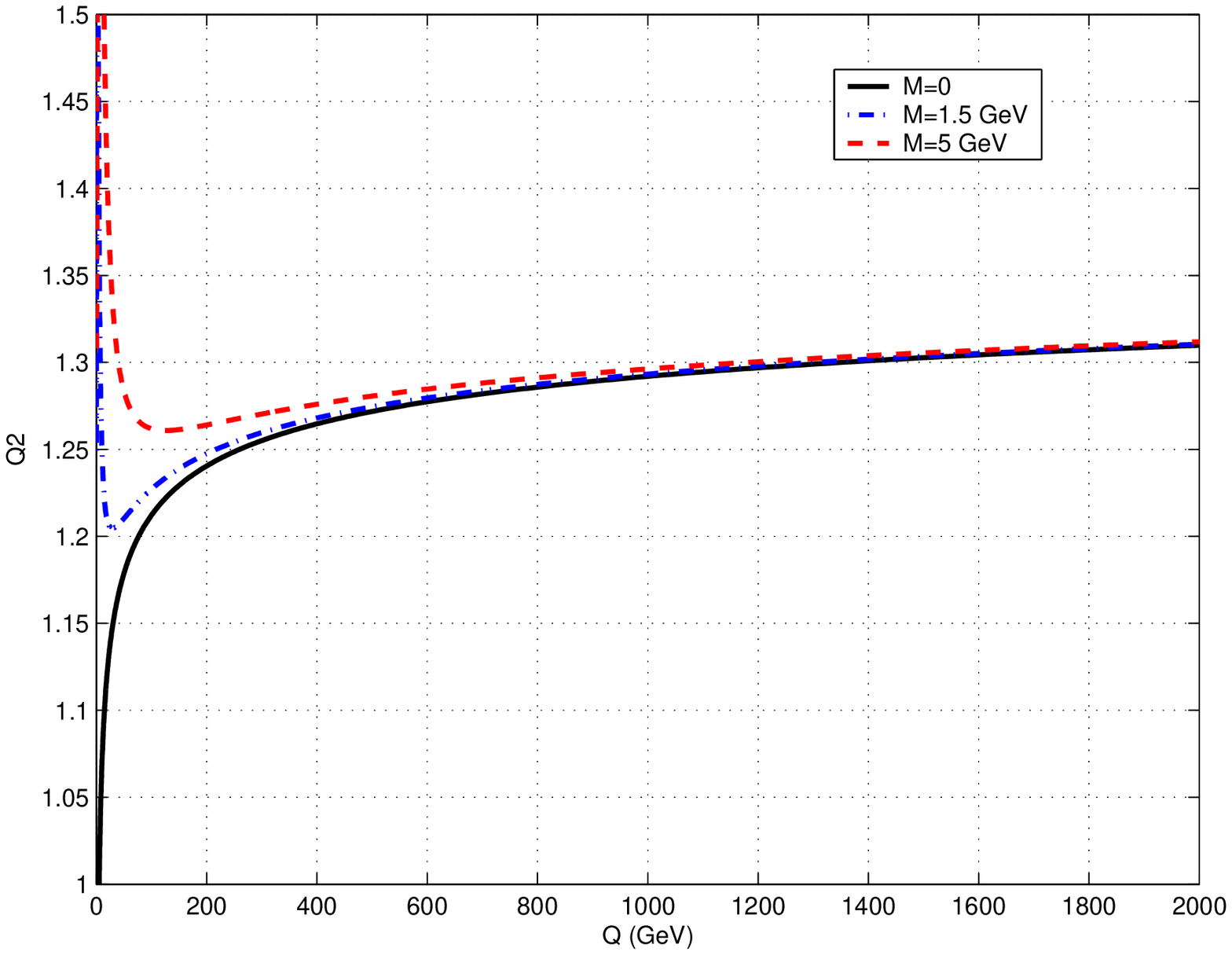, height=6.5truecm,width=7.5truecm}
\epsfig{file=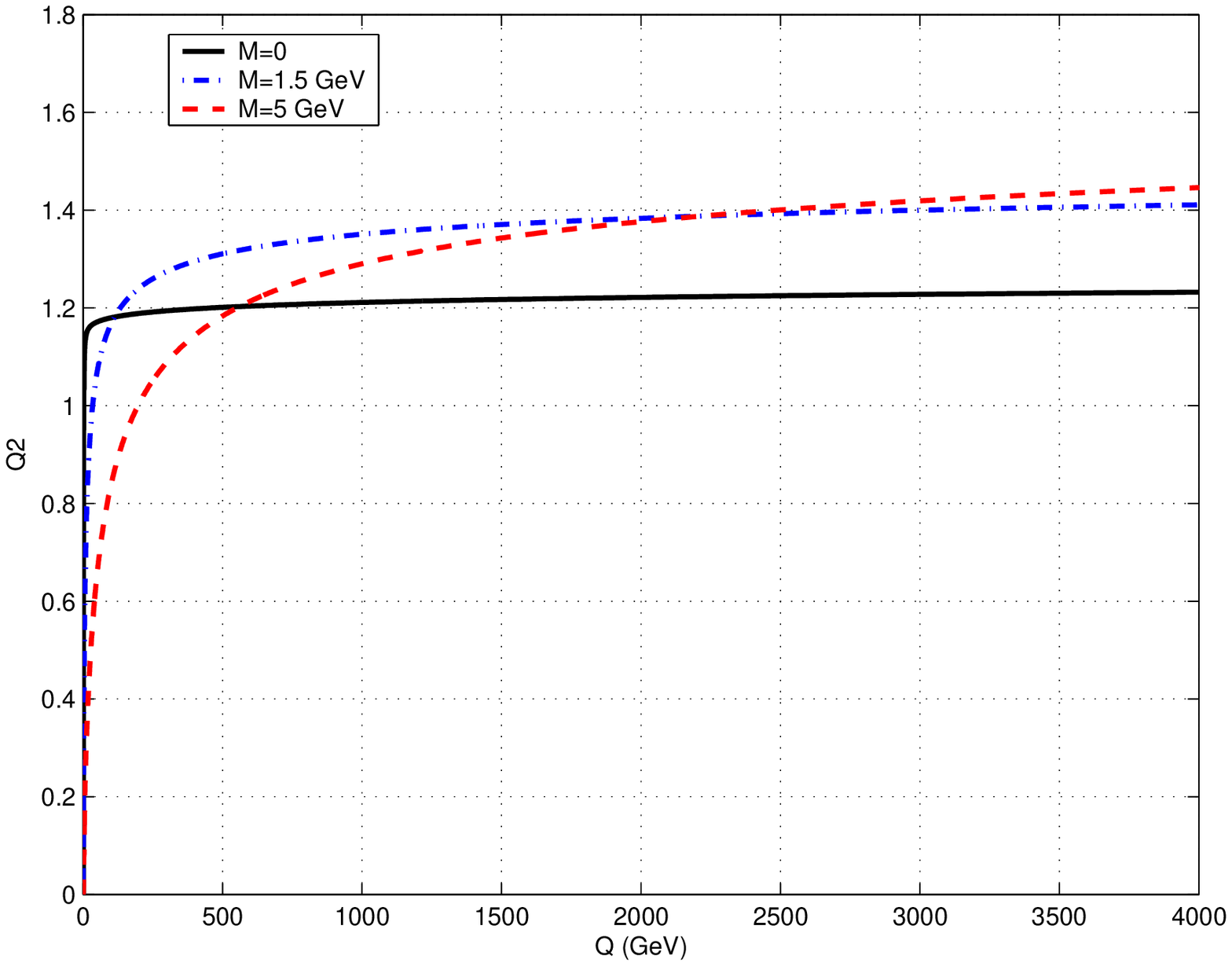, height=6.5truecm,width=7.5truecm}
\caption{\label{fig:analitnumer} Analytical (left) versus numerical (right) 
solution of equation (\ref{eq:N2Qh}) for the second multiplicity correlator $Q_2$
defined in (\ref{eq:G2Q2}).}  
\end{center}
\end{figure}

In Fig.\,\ref{fig:analitnumer} we display the analytical solution together with the numerical 
solution of (\ref{eq:N2Qh}) for the second multiplicity correlator $Q_2$ defined in (\ref{eq:G2Q2}). 
As it can be seen, when the mass of the leading heavy quark increases, 
the approximated analytical correlator becomes slightly stronger. However, because of forward gluon 
suppression taken into account by the integrated function $\epsilon_1(\tilde Y,L_m)$ in the leading DL 
contribution of (\ref{eq:N2Qh}), such a behaviour cannot
be trusted for lower virtualities than 
few thousands of GeV. That is why, even if the shape of the analytical solution may be 
correct for $Q\gtrsim100$ GeV, we should only trust the shape and normalization of the 
numerical solution in a much wider energy range $Q\gtrsim40$ GeV in view of realistic QCD predictions.

\bibliographystyle{plain}

\bibliography{mybib}

\end{document}